# A bulge test based methodology for characterizing ultra-thin buckled membranes

Salman Shafqat[1], Olaf van der Sluis[1, 2], Marc Geers[1], and Johan Hoefnagels[1, †]

[1] Department of Mechanical Engineering, Eindhoven University of Technology, 5600 MB Eindhoven, the Netherlands;

s.shafqat@tue.nl; o.v.d.sluis@tue.nl; m.g.d.geers@tue.nl; j.p.m.hoefnagels@tue.nl

[2] Philips Research, High Tech Campus 4, 5654 AE Eindhoven, the Netherlands

[†] Corresponding author: j.p.m.hoefnagels@tue.nl; Tel.: +31-40-2475894

## Abstract

Buckled membranes become ever more important with further miniaturization and development of ultra-thin film based systems. It is well established that the bulge test method, generally considered the gold standard for characterizing freestanding thin films, is unsuited to characterize buckled membranes, because of compressive residual stresses and a negligible out-of-plane bending stiffness. When pressurized, buckled membranes immediately start entering the ripple regime, but they typically plastically deform or fracture before reaching the cylindrical regime. In this paper the bulge test method is extended to enable characterization of buckled freestanding ultra-thin membranes in the ripple regime. In a combined experimental-numerical approach, the advanced technique of digital height correlation was first extended towards the sub-micron scale, to enable measurement of the highly varying local 3D strain and curvature fields on top of a single ripple in a total region of interest as small as ~25 µm. Subsequently, a finite element (FE) model was set up to analyze the post-buckled membrane under pressure loading. In the seemingly complex ripple configuration, a suitable combination of local region of interest and pressure range was identified for which the stress-strain state can be extracted from the local strain and curvature fields. This enables the extraction of both the Young's modulus and Poisson's ratio from a single bulge sample, contrary to the conventional bulge test method. Virtual experiments demonstrate the feasibility of the approach, while real proof of principle of the method was demonstrated for fragile specimens with rather narrow (~25 µm) ripples.

## 1. Introduction

The bulge test methodology has become the standard technique for mechanical characterization of thin films [1] especially for freestanding membranes. This is due to (1) the possibility of precise sample processing facilitated by recent developments in micro-fabrication technology; (2) the need for minimal sample handling, which is especially challenging at small scales; (3) the relatively simple data processing for determining the membrane stress and strain values, needed to extract the mechanical properties. The method essentially involves fixing a freestanding membrane over a small window opening and applying a known pressure to it, while measuring the resulting membrane deflection (or curvature). Various models have been developed, based on the sample geometry, to convert the pressure-deflection data to the a (elasto-plastic) stress-strain curve, which is used to determine mechanical properties such as Young's modulus, Poisson's ratio, residual stresses, and plasticity parameters [1–3]. During the last 30 years ample research has been devoted to improve the accuracy of the bulge test by studying the underlying assumptions such as the influence of bending stiffness [4] and initial conditions e.g. initial film thickness and residual stress [1,5,6]. Among the different varieties of the bulge test method, the plane-strain bulge test is most popular [7], where it was shown that for rectangular membranes with in-plane aspect ratio larger than 4, the stress state in the center of the membrane reduces to a plane strain condition. This means that the stress and the strain are given by [4]:

$$\kappa_{tt} = \frac{2\delta}{a^2+\delta^2}, \qquad (1)$$

$$\sigma_{tt} = \frac{P}{h\,\kappa_{tt}}, \qquad (2)$$

$$\epsilon_{tt} = \frac{1}{a\kappa_{tt}}\sin^{-1}(a\kappa_{tt}) - 1, \qquad (3)$$

where $\kappa_{tt}$ is the curvature in the transverse direction, $a$ is half of the width of the membrane, $\delta$ is the deflection of the apex of the membrane, $P$ is the applied pressure and $h$ is the membrane thickness. $\sigma_{tt}$ and $\epsilon_{tt}$ denote the normal stress and strain in the transverse direction, respectively. As a consequence of the plane strain condition in the center of the rectangular membrane, the transverse stress and the transverse strain can be related by the following constitutive equation:

$$\sigma_{tt} = \left(\frac{E}{1-\nu^2}\right)\epsilon_{tt}, \qquad (4)$$

where $\left(\frac{E}{1-\nu^2}\right)$ is the plane strain modulus. To extract the Young's modulus ($E$) and the Poisson's ratio ($\nu$) separately, an additional test needs to be performed, e.g., a bulge test on a circular or square membrane for which a different stress-strain equation holds, resulting in the biaxial modulus $\left(\frac{E}{1-\nu}\right)$ [2,8].

However, such freestanding membranes often buckle as a result of processing induced (compressive) residual stresses in combination with their small out-of-plane bending stiffness, particularly for ultra-thin membranes. In some cases, the buckling is exploited as a functional part in devices, e.g., in bi-stable micro actuators [9–11]. Moreover, the buckling phenomenon in freestanding thin membranes has gained a lot of attention in Micro Solid Oxide Fuel Cells (μSOFC), where stacks of freestanding membranes serving as electrodes or solid electrolytes, are often buckled as a result of the processing. While initially buckling was considered an issue [12], recent literature suggests that it can actually be beneficial to have these membranes in a buckled state to enhance their functional properties. It has been shown that buckled membranes are mechanically more stable at elevated temperatures, i.e. lower thermomechanical tensile stresses develop compared to a 'flat' membrane, often having significant tensile stresses already at room temperature [13], [14]. Mechanical models have been developed to exploit the behavior of such μSOFC membranes in the post-buckling regime and consequently expand the design space into the low-stress post-buckling regime [12,13]. Recently, controlled buckling patterns in μSOFC solid electrolyte membranes (Figure 1a) using 'strain engineering' have been employed to demonstrate local tuning of ionic conductivity of the electrolyte as an alternative of solid solution doping [15]. Furthermore, in the exciting field of graphene, where buckles and ripples are intrinsically present in the suspended configuration (Figure 1b), these phenomena are receiving considerable attention [16] to be exploited in various applications [17], such as improved hydrogen absorption on a rippled graphene surface (due to local curvature), for future efficient hydrogen-based fuel cells [18].

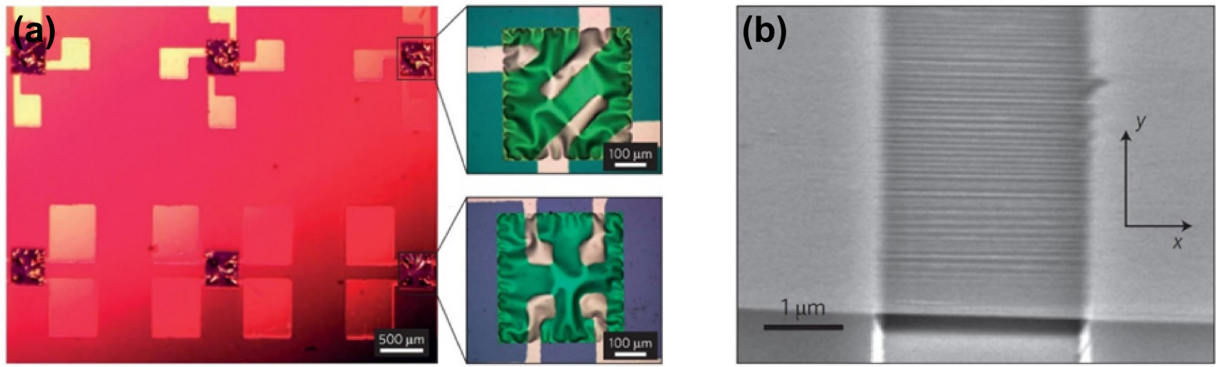

*Figure 1: Examples of applications of buckled membranes: (a) design and placement of electrodes to locally strain-engineered ionic transport of the solid electrolyte freestanding membrane for µSOFC application [15] (reproduced with permission), (b) SEM micrograph showing ripples in a bilayer suspended graphene membrane [16] (reproduced with permission).*

The presence of such buckling patterns in test samples, as shown in Figure 2, prevents the application of the conventional bulge test. All available literature to date confirms that the conventional bulge test methodology cannot be used to characterize buckled membranes since even at high pressures, buckling patterns typically do not disappear in the membranes and some stress components near the edge of the membrane stay compressive [2,19]. Only in very few cases, the samples can be pressurized, beyond the point where the buckling pattern completely disappears, where the bulge equations (1)-(4) might apply [2,19]. Alternatively, the sample may deform plastically before entering the cylindrical regime. Therefore, such buckled samples are typically discarded and the processing needs to be modified to prevent the buckles to occur, in order to mechanically characterize the membranes accurately. Such processing modifications can be time consuming, costly and sometimes even infeasible or undesired. Moreover, any processing change could influence the actual properties to be determined.

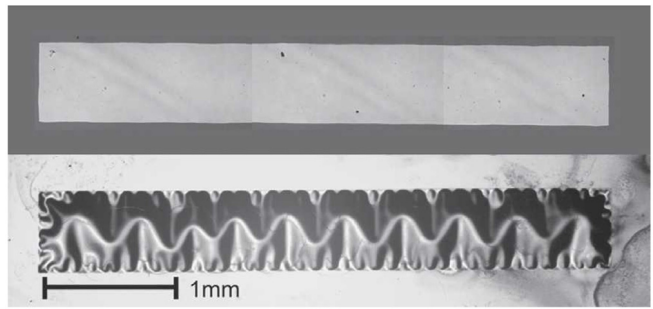

*Figure 2: (Top) Typical 'flat' freestanding bulge test membrane. (Bottom) Buckled membrane meandering/telephone cord type pattern* [19] *(reproduced with permission).*

Clearly there is a need for a convenient characterization methodology to determine the material properties of the buckled samples in their original (buckled) state.

This paper introduces a characterization methodology for testing buckled samples, which builds on the bulge test theory and thus exploits its aforementioned advantages. The approach adopted here is to numerically model the bulge test-like pressure loading of the buckled membrane in the rippled regime, to which the meandering pattern (see Figure 2 bottom) starts transitioning as soon as even minute pressure is applied [20], to understand the mechanics and provide relations for the relevant membrane stress and strain components. To relate the stresses and strains using simple constitutive equations for extracting the material properties, regions of interest (ROI) with simplified stress states are identified and explored. Furthermore, to accurately measure the complex non-uniform three-dimensional displacement field of the buckled membranes, recent advances in bulge test methodology involving integration of Global Digital Image Correlation (GDIC) with conventional bulge test theory to [8] are exploited.

## 2. Methodology

### 2.1 Digital height correlation based bulge test

In conventional bulge test theory, the stresses and the strains obtained using equation (2) and (3) are based on the assumption that they are homogeneous over the membrane, i.e. an infinitely long cylinder or full sphere is assumed. This assumption does not hold anymore when the bending effects at the boundaries play a significant role for films with a relatively large thickness [8]. Inhomogeneous fields are also expected in the case of buckled membranes, even for a very small thickness, and in the pressure loaded case. This challenge is addressed by adopting a recently developed Digital Height Correlation (DHC) based bulge test technique [8], to capture the non-uniform 3D displacement fields.

The fundamental concept underlying this extension is to apply digital image correlation on the topographical (height) maps of subsequent load increments (pressure increments), resulting in corresponding displacement fields. Curvature fields can subsequently be computed from the displacement fields. Based on the sample geometry, relations such as equation (2) can be used locally to obtain stress from curvature data, while the local strain can be obtained directly from the displacement fields. Therefore, the key assumption of a uniform cylindrical shape and uniform deformation adopted in the conventional bulge test methodology does not need to be fulfilled. Consequently, more accurate, local stress and strain fields can be obtained.

Digital Height Correlation is a variant of Global Digital Image Correlation (GDIC) and it is based on the conservation of height, instead of brightness, between a (topographical) reference image $f(\vec{x})$ and a corresponding deformed image $g(\vec{x})$, where $\vec{x}$ is the in-plane position vector, i.e. $\vec{x} = x\vec{e}_x + y\vec{e}_y$, with $\vec{e}_x$ and $\vec{e}_y$ denoting the Cartesian unit vectors. The conservation of height is written as:

$$f(\vec{x}) = g\big(\vec{x} + u_x(\vec{x})\vec{e}_x + u_y(\vec{x})\vec{e}_y\big) - u_z(\vec{x}), \qquad (5)$$

where $u_x$ and $u_y$ are the in-plane displacement components in $x$ and $y$ direction respectively, while $u_z$ is the out-of-plane displacement component. The image residual $r(\vec{x})$ is defined as:

$$r(\vec{x}) = f(\vec{x}) - g\big(\vec{x} + u_x(\vec{x})\vec{e}_x + u_y(\vec{x})\vec{e}_y\big) - u_z(\vec{x}) + n_0(\vec{x}), \qquad (6)$$

$$\approx g(\vec{x}) + \big(\vec{\nabla}g \cdot \vec{e}_x\big)u_x(\vec{x}) + u_y(\vec{x}) + \big(\vec{\nabla}g \cdot \vec{e}_y\big)u_y(\vec{x}) - u_z(\vec{x}) + n_0(\vec{x}), \qquad (7)$$

where $n_0$ is the image noise and $\vec{\nabla}g$ is the gradient of the deformed image $g$. The square of the image residual is minimized over the region of interest in the GDIC algorithm,

$$\gamma^2 = \int_{ROI} r(\vec{x})^2 d\vec{x}, \qquad (8)$$

where $\gamma$ is the global residual. The 3D displacement vector is given by:

$$\vec{u}(\vec{x}) = u_x(\vec{x})\vec{e}_x + u_y(\vec{x})\vec{e}_y + u_z(\vec{x})\vec{e}_z. \qquad (9)$$

To make this a well-posed optimization problem, the displacement field is parameterized as a sum of basis functions $\varphi_i(\vec{x})$ that act over the ROI and weighted by a discrete set of degrees of freedom $\lambda_i$. The choice of the basis and shape functions is based on the expected deformation complexity. It should be noted that different shape functions may be required in the $x, y$ and $z$ directions to capture the deformation. It has been shown that the shape of square and rectangular bulged membranes are well described by polynomial functions [21].

$$\vec{u}(\vec{x}) = \sum_i^l \lambda_i \varphi_i(\vec{x}) \vec{e}_x + \sum_i^m \lambda_i \varphi_i(\vec{x}) \vec{e}_y + \sum_i^n \lambda_i \varphi_i(\vec{x}) \vec{e}_z, \tag{10}$$

where the basis functions $\varphi_i$ are chosen here to be polynomial functions, given by:

$$\varphi_i = x^\alpha y^\beta \tag{11}$$

Subsequently, the strain fields can be computed from the extracted displacement fields using an appropriate strain definition. To determine the relevant stress components, the curvature field is obtained with the same DHC measurement. The curvature tensor $\boldsymbol{\kappa}$ is determined by taking the spatial gradient of the outward normal vector $\vec{n}$ field as:

$$\boldsymbol{\kappa}(\vec{x}) = \vec{\nabla} \otimes \vec{n}(\vec{x}), \tag{12}$$

where the outward normal vector is the normalized gradient of the position field $z(\vec{x})$, as given by:

$$\vec{n}(\vec{x}) = \frac{\vec{\nabla} z(\vec{x})}{||\vec{\nabla} z(\vec{x})||}. \tag{13}$$

The curvature fields in transverse direction is given by:

$$\kappa_{tt}(\vec{x}) = \vec{t}_x(\vec{x}) \cdot \boldsymbol{\kappa}(\vec{x}) \cdot \vec{t}_x(\vec{x}), \tag{14}$$

where $\vec{t}_x$ is a vector tangent to the membrane surface along the $x$ direction. These curvature fields will be used to determine the local stress components. For instance, for a local region with a plane strain state (in the middle of a non-buckled rectangular membrane), the curvature can be related to the hoop stress by:

$$\sigma_{tt} = \frac{P}{t\kappa_{tt}}, \tag{15}$$

where $P$ is the applied (uniform) pressure and $t$ is the thickness of the membrane.

## 2.2 Experimental setup and procedure

A custom-made, gaseous pressure medium based bulge test setup (similar to the setup reported in Ref. [22]) is used (see Figure 3b) for the experiments. The setup consists of a sample holder block attached to a pressurized N₂ reservoir through a pressure regulator. A commercially available pressure regulator (MFCS-EZ by Fluigent) was used with a range of $0 - 200\ kPa$ and a resolution of $6\ Pa$ (0.03% of full range). The pressure regulator and consequently the bulge test setup has a fast response and settling time, which is an advantage over liquid based setups. While liquid pressure-medium setups reach higher pressures, as needed for testing stiff (thick or narrow) films or plates, they can suffer adversely from pressure buildup problems in case of a minute leakage or the presence of gas bubbles in the pressure medium. This is usually not a problem in gaseous pressure-medium setups, since any drop in pressure due to leakages is directly compensated through the connected large reservoir. Furthermore, the setup is less sensitive to pressure changes due to ambient temperature variations compared to liquid based setups.

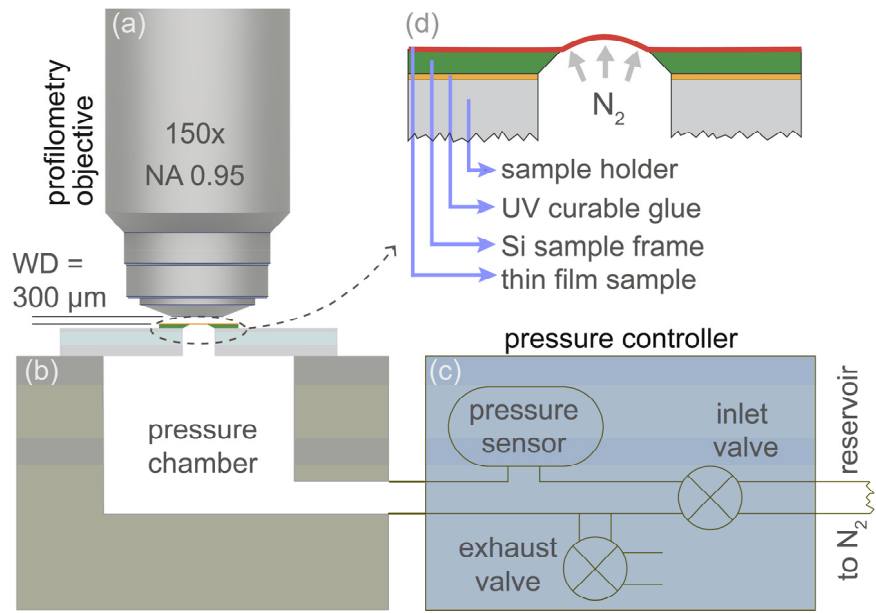

*Figure 3: Schematic of experimental setup for performing bulge test experiments: (a) high-resolution optical profilometry object, (b) bulge test setup, (c) pressure regulator to control pressure loading with regulator N₂ supplied with through a cylinder reservoir, (d) magnified view of cross-section of a typical bulge sample with a thin film deposited in a silicon frame with an etched window in the center. The sample is fixed by gluing it onto the bulge tester sample holder.*

The setup is placed under a commercially available optical profilometer, Sensofar Plµ 2300. The profilometer is used in confocal mode to obtain full-field topographical images. The highest magnification lens, with a magnification of $150x$ and Numerical Aperture (N.A.) of 0.95 is used to obtain high resolution images of the narrow ripples which are $20 - 30\ \mu m$ wide. The resulting field of view is $84x63\ \mu m^2$ with in-plane spatial sampling (pixel size) of $0.11\ \mu m$, while the effective height resolution obtained is in the range of $25\ nm$.

Rectangular bulge test samples used for the proof of principle experiment were fabricated by deposition of a (proprietary) multi-layered stack consisting of transition metals and oxides on a monocrystalline 650 $\mu$m thick silicon wafer. The freestanding window was created by wet etching from the back side of the wafer with a $1x5\ mm^2$ window. The sample has a total thickness of $64\ nm$ and (volume averaged) Young's modulus (estimated by volume averaging using the rule of mixtures) of $217\ GPa$ and Poisson's ratio of 0.35. In general, application of DIC requires a good pattern providing sufficient image contrast. In case of digital height correlation this contrast is achieved by local differences in height on the sample surface. Since the native surface roughness of the samples is in the order of a few nanometers, below the resolution of the profilometer, a 'height' pattern is applied. For this purpose, 500 $nm$ mono-dispersed polystyrene microspheres (by micromod®) were used. The microspheres were applied using the drop-casting method. The particles are provided in a dense suspension and are further diluted (by a dilution factor of ~40) in ethanol to achieve the required particle density on the sample surface (see Figure 4). A relatively dense pattern is required to capture the expected inhomogeneous displacement fields. Since the particles do not form a continuous layer, adhering to the sample surface (upon contact, without the need for an adhesive) as single particles or homogeneously distributed aggregates composed of few particles, their influence on the mechanics of the membrane is assumed to be negligible. Moreover, during the pressure loading step, the particles adhere well to the sample and no pattern change or degradation is observed, which is important for reliable application of DIC.

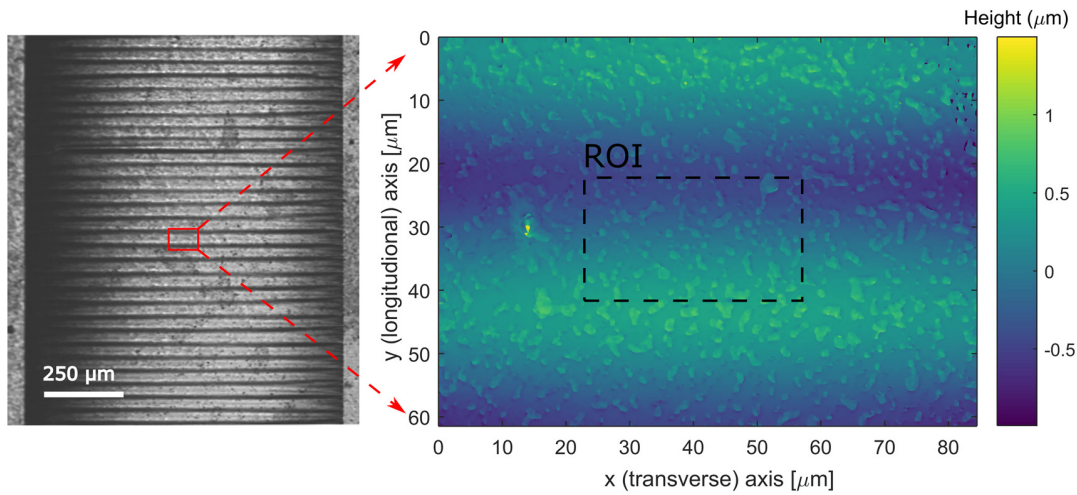

*Figure 4: Images of ripples in a pressurized membrane with, optical image (left) and topographical image of an area imaged during the experiment, marked in the optical image by the red rectangle (right). The region of interest marked by the rectangle with black dashed lines is selected for DHC. The roughness pattern visible in the topographical image consists of clusters of 500 nm polystyrene microspheres, applied as a DHC pattern.*

## 2.3 Numerical modelling

A non-linear FE model was set up to simulate the pressure loading of the test sample using the commercial FE program, MSC Marc/Mentat®. Since in the test specimens the meander profile starts transitioning to a ripple profile as soon as a slight pressure is applied [20] [see Figure 6], the analysis is focused on ripple and the cylindrical regime. As the substrate is many orders of magnitude stiffer than the film, it is modelled with rigid boundary conditions at the edges of the film. The rectangular membrane was meshed uniformly with quadrilateral 4-node thick-shell elements having three translational and three rotational degrees of freedom at each node. A mesh convergence study was performed and it was found that the solution becomes mesh independent at a size of 80×400 elements (used as the mesh size for the model).

Four batches of multi-layer test samples were available for testing, with varying material stack thickness ratios and hence different overall membrane thickness and volume averaged Young's modulus. The focus here is on developing a methodology for obtaining thickness-averaged mechanical properties of the buckled membranes. This is also the case for conventional bulge test which also provides thickness-averaged data only. Therefore, the Young's modulus and Poisson's ratio in the FE model were set to $153.45\ GPa$ and $0.351$, respectively, based on a batch of samples planned for fabrication but was never produced. Note, however, that the exact properties used for the FE model are unimportant for the analysis and conclusion made and the model serves as a general platform for virtual experiments. The dimensions of the membrane in the FE model were taken as $1 \times 5\ mm^2$ in accordance with the size of the test samples, while the membrane thickness was set to a value of $50\ nm$ based on the median thickness batch.

To induce buckling in the membrane in-plane compressive load is required. The residual compressive stress was simulated through a thermal loading step resulting in an in-plane stress applied by the substrate (frame) on the freestanding membrane. This is caused by higher contraction of the substrate w.r.t the membrane on cooling down from a high temperature. Since the substrate was modelled by rigid membrane boundaries, the difference in coefficient of thermal expansion ($\alpha$) of the substrate and the membrane ($-2.6 \times 10-6\ °C^{-1}$) was assigned to the membrane, to simulate this effect.

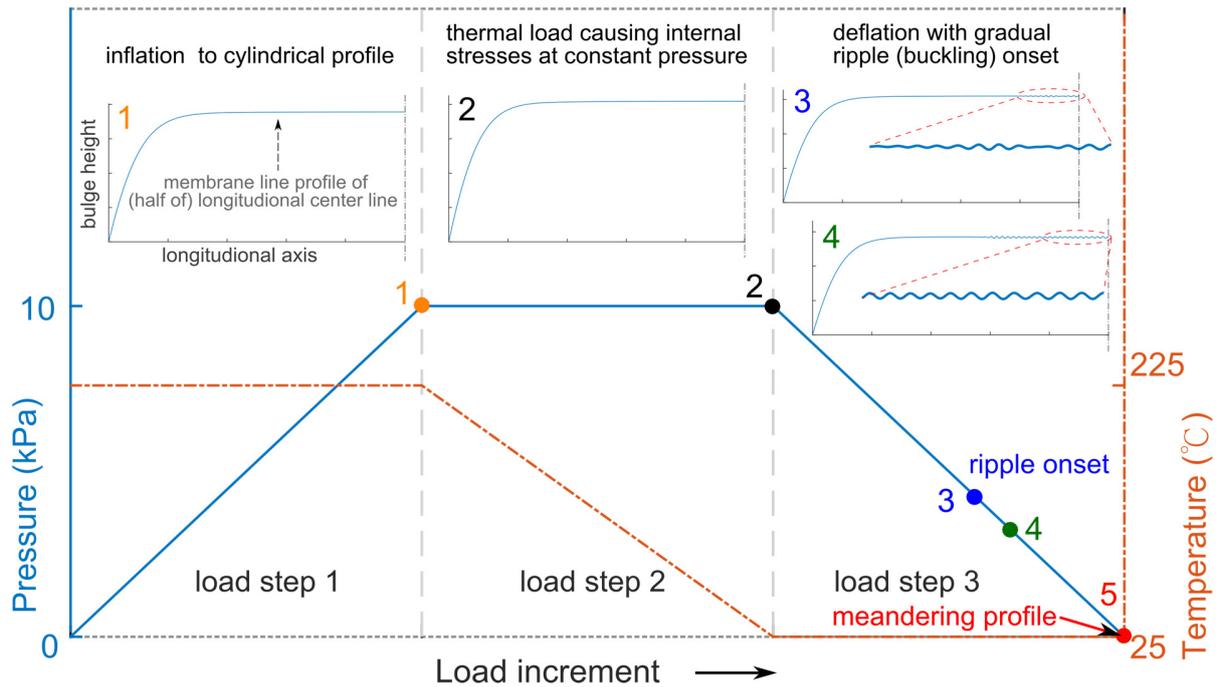

*Figure 5: Three-step thermo-mechanical loading procedure for gradual onset of buckling, initially as a ripple profile, finally settling into a meandering profile at the last load increment. In load step 1, the pressure is increased while temperature is kept constant to result in an inflated cylindrical profile. Subsequently, in load step 2, the temperature is decreased while the pressure is kept constant, resulting in thermal stresses that (gradually) manifest themselves as increasing residual compressive stress while the pressure is reduced (as the temperature is kept constant) in load step 3, triggering the onset of ripples. Note that the top sub-figures represent membrane defection for the corresponding load increment with a line profile of half of the longitudinal center line.*

In order to avoid typical numerical instabilities at the bifurcation point (i.e. suddenly going from a flat to meandering shape in a single increment), a method used to model strongly buckled square membranes from Ref. [23] was adopted here. This involves bypassing the bifurcation point by adopting the three-step loading procedure illustrated in Figure 5. First the membrane is bulged by application of a pressure load (Figure 6a$_3$). Then, compressive residual stresses are induced in the bulged membrane by applying the thermal loading step. In the final step the pressures is gradually decreased to let the membrane slowly settle down into a rippled profile (Figure 6a$_2$). If the pressure is completely removed the membrane transitions from a rippled to a meandering configuration (Figure 6a$_1$). Using this method, in addition to bypassing the bifurcation point, the strong and sudden geometrical nonlinearities expected at the transition from planar to meandering configuration are avoided by gradually settling from the cylindrical to the rippled configuration.

# 3. Numerical analysis

## 3.1 Simulation results

Using the three step loading procedure, explained in the previous section, the numerical model adequately captures the three different regimes seen in the experiments, i.e. the rippled, meandering regime, and the cylindrical regime. Moreover, the evolution of the rippled regime, with increasing pressure as well as the transitions between the different regimes seems are well captured. Comparison with experimentally observed regimes shown in Figure 6 provides a qualitative validation of the model.

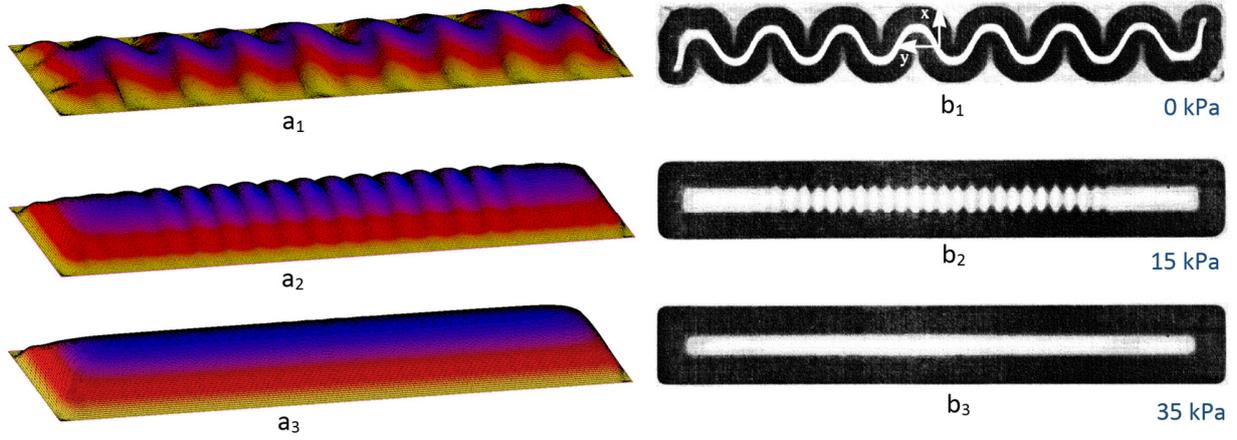

*Figure 6: Comparison between numerically modelled (a) and experimentally observed [24] (b) buckling regimes: meandering regime ($a_1, b_1$), rippled regime ($a_2, b_2$) and cylindrical regime ($a_3, b_3$), with membrane size 1x5 $mm^2$ (a) and 0.65x4.5 $mm^2$ (b). Note that similar buckling (meandering and rippled) regimes as in Ref. [24] are observed in our samples (see Figure 4), however, since cylindrical regime is not reached, the images are not shown here.*

In order to also quantitatively validate the FE model, the Energy Minimization Method (EMM) based model reported by Kramer et al. ([24]) is exploited here. This model was developed to describe the rippled regime for a similar rectangular membrane with width $a$ and thickness $h$, as discussed here. The model is able to predict the reduced ripple wavelength $\bar{\lambda} = \frac{\lambda}{a}$, the reduced peak-to-peak amplitude $\Delta\bar{w} = \frac{\Delta w}{h}$ and the reduced ripple free amplitude $\bar{w}_{ps} = \frac{w_{ps}}{h}$ (see Figure 7), as a function of the reduced prestrain $\bar{\epsilon}_0 = \frac{\epsilon a^2}{h^2}$ and the reduced pressure $\bar{p} = \frac{p(1-\nu^2)a^4}{Eh^4}$.

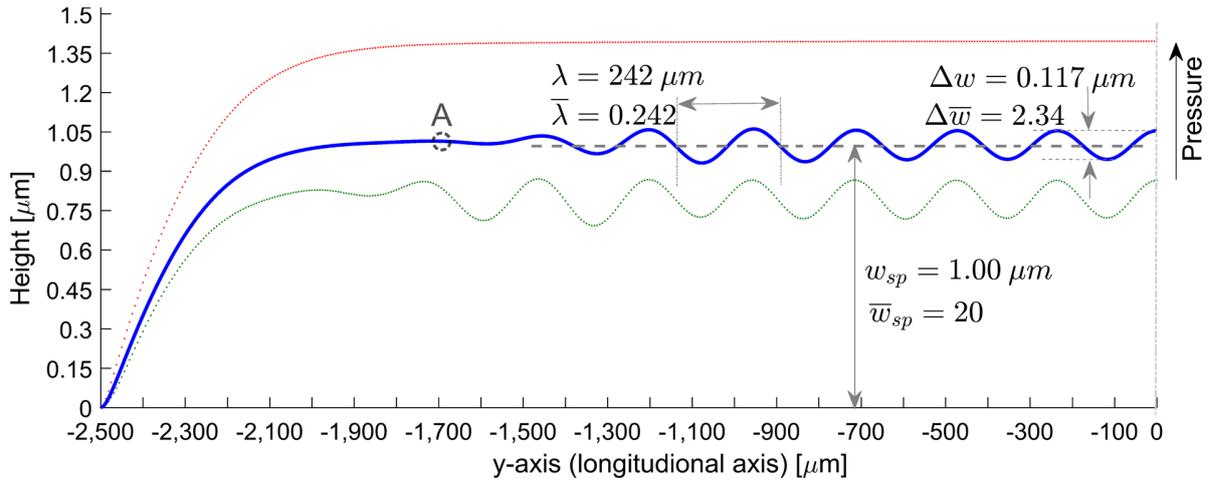

*Figure 7: Centre line profile of (half of) the membrane along the longitudinal direction, extracted from FE simulation at increasing pressure values loads, with the middle profile (in blue) corresponding to $\bar{\epsilon}_0$= -500 and $\bar{p}$ = 7.48×10⁴ displaying relevant parameters. The three profiles display the evolution of the buckling pattern with increasing pressure.*

For realistic values of the applied pressure and residual stress (due to thermal loading) the results for our test specimens lie significantly outside the boundaries of the plotted results presented in Ref. [25], due to their very small thickness. In order to validate the FE model, smaller temperature-load and pressure values are applied to enable a comparison with the EMM results. Based on the values of $E$, $\nu$ and the coefficient of thermal expansion ($\alpha$), used in the FE model, a temperature difference of $0.347\ °C$ and a pressure load of $0.0818\ Pa$ is calculated, which corresponds to a reduced strain, $\bar{\epsilon}_{0test}$ of $-500$ and a reduced pressure $\bar{p}_{test}$ of $7.4811 \times 10^4$, respectively, thus bringing the $\bar{\epsilon}_0$ and $\bar{p}$ values within the bounds of the reported EMM results available (in Ref. [25]). A line profile along the longitudinal axis in the middle of the membrane from the FE simulations is displayed in Figure 7 with the calculated reduced parameters which are given in Table 1, along with the corresponding parameters from the EMM results at the same $\bar{\epsilon}_0$ and $\bar{p}$ values.

Table 1: Comparison of $\bar{\lambda}$, $\bar{w}_{ps}$ and $\Delta\bar{w}$ predicted by EMM [25] and by the present FE analysis

|  | FE | EMM |
|---|---|---|
| $\bar{\lambda}$ | 0.242 ± 0.01 | 0.250 ±0.005 |
| $\bar{w}_{ps}$ | 20 ± 0.4 | 19.7 ± 0.2 |
| $\Delta\bar{w}$ | 2.3 ± 0.4 | 2.6 ± 0.1 |

As can be seen in Table 1, the FE and EMM results agree for all three parameters within readout error from the FE and EMM results. However, the deviation for $\Delta\bar{w}$ is significant. On the one hand, the higher variation for $\Delta\bar{w}$ can partially be attributed to a higher readout error from the EMM plots, reflecting in the error bar in Table 1. On the other hand, the peak-to-peak amplitude varies over the length of the line profile in the FE results (see Figure 7), thus an average value of a relatively small magnitude is taken thereby, possibly contributing to the relatively high deviation. Based on the adequate agreement for all three values, FE model is considered valid.

## 3.2  Numerical analysis of a suitable regions of interest:

As the model adequately captures the mechanics of the pressure loaded buckled membrane, it is well suited to be used as a numerical framework to test the methodology developed here, i.e.

serving as a virtual experiment. Various ROIs are analyzed, where, with suitable modifications, bulge test analysis can be applied. The criteria for choosing a suitable ROI are:

1. The presence of a simplified stress or strain state, e.g. plane-strain
2. Membrane stresses can be conveniently calculated using an analytical equation, e.g. Eq. (2)
3. Membrane strains (using DHC) can be accurately determined

While the cylindrical regime may seem promising from the analysis point of view as classical simple plane strain bulge equations may be applicable there, most initially-buckled membranes (of various types) fracture before reaching the cylindrical regime, while those membranes that can sustain a high enough pressure typically go into plasticity before entering the cylindrical regime. Alternatively, the meandering regimes is not interesting form a practical point of view, as it is only accessible at minute pressures [20]. Since the rippled regime exists throughout almost the whole pressure loading cycle, only the rippled regime is here considered for further analysis. Therefore, the analysis must be performed in the ripple regime.

There are multiple reason for choosing the ROI along the longitudinal center line of the membrane. First, it is observed in the FE simulations and the experiments that the magnitude of the ripples close to the longitudinal edge is lower than that of the same ripples in the center of member, see Figure 8. This is due to the boundary constraints provided by the edge (on the edge ripples can, of course, not form). Second, considering the third criterion of accurate determination of the membrane strains using DHC keeping the ROI always in the field of view is important. Due to symmetric deformation, an ROI in the center of the membrane experiences the least rigid body motion, while closer to the membrane edge the large out of plane rotations result in large rigid body motion. Furthermore, these large out-of-plane rotations can affect the image residual, due to the change in effective viewing angle of the pattern, as discussed below. Finally, the boundary conditions at the edge are never as perfect as assumed in a Finite Elements simulation, especially when a dry etch is used to free the membrane, therefore, it is best to do the analysis far away from the edge, where these local boundary effects are negligible due to the well-known Saint-Venant's principle in solid mechanics. Therefore, the most suitable ROI is identified as an ROI along the longitudinal center line far away from the edges.

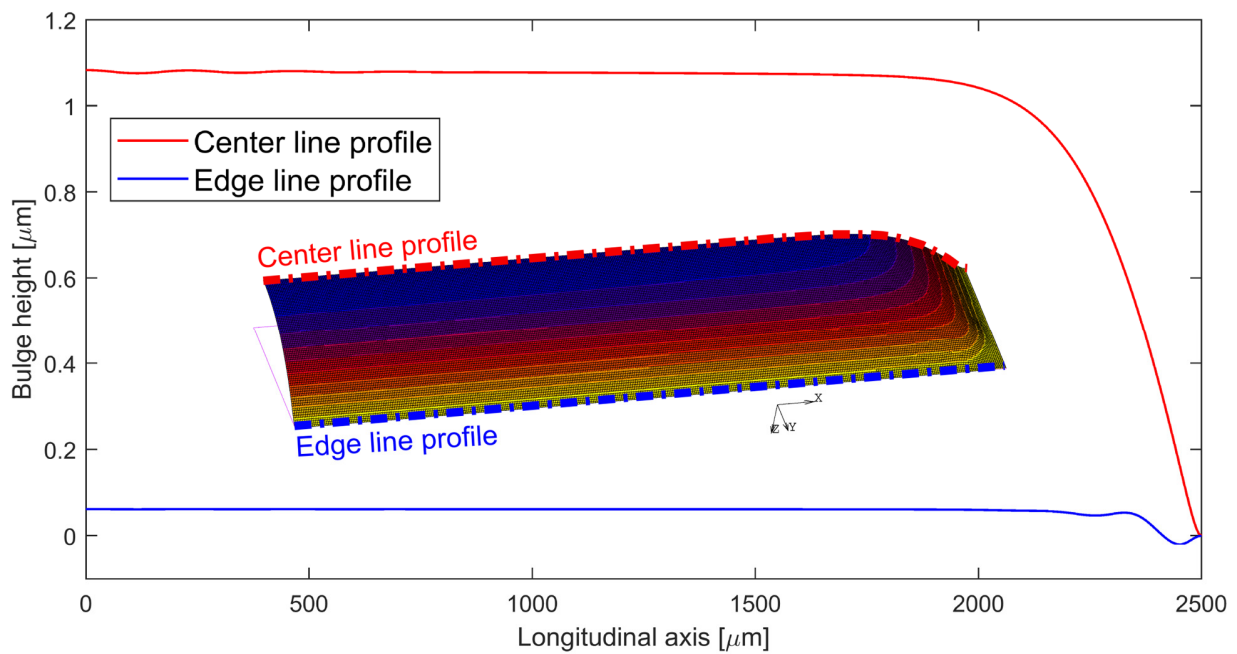

*Figure 8: Comparison of simulated center line profile and edge line profile (calculated at the first free node from the edge), at the first FEM increment that shows ripples (i.e. point 3 in Figure 5).*

It can be assumed that the membrane is in a state of plane stress with respect to the thickness direction, since the thickness of the membrane is very small relative to the other two dimensions (with free contraction in the thickness direction). Indeed, the stress in the thickness direction due to the applied pressure is negligible compared to the in-plane stress. Furthermore, isotropic material behavior is assumed (which is common for thin films produced with thin film deposition techniques). Therefore, the isotropic linear elastic plane stress equations (Eq. (16) and (17)) apply for the present analysis.

$$\epsilon_{yy} = \frac{1}{E}\sigma_{yy} - \nu\frac{1}{E}\sigma_{xx}, \tag{16}$$

$$\epsilon_{xx} = -\nu\frac{1}{E}\sigma_{yy} + \frac{1}{E}\sigma_{xx}, \tag{17}$$

where $\epsilon_{yy}$ and $\epsilon_{xx}$ are the normal membrane strains in the longitudinal and transverse direction, respectively, while, $\sigma_{yy}$ and $\sigma_{xx}$ are the membrane stresses in the longitudinal and transverse direction respectively.

The FE analysis shows that in the cylindrical regime the magnitude of the transverse stress $\sigma_{xx}$ is much larger than the longitudinal stress $\sigma_{yy}$ as expected for a rectangular geometry. In the last loading step, as the pressure is being reduced, at a certain stage (labelled point 3 in Figure 5), the longitudinal stress becomes compressive while the transverse stress is still tensile. As soon as the stress state in the longitudinal direction becomes compressive, the membrane releases the compressive stress by buckling, in the form of a rippled pattern. Given its small thickness and consequently negligible bending stiffness, the membrane does not possess the ability to support a compressive stress, which is therefore released in a buckling pattern. This phenomenon is illustrated in Figure 9. After the emergence of the ripples, as the pressure is further decreased, the longitudinal stress remains almost zero. The transverse stress however is still tensile and it keeps on decreasing with decreasing applied pressure. The small longitudinal stress varies over the width of a ripple due to the bending induced stress, and is therefore most compressive at the valley of the ripple while being least compressive at the peak of the ripple. At the crossover point in the middle of a ripple where the curvature is zero (marked with point C in Figure 11a) however, where bending effects do not contribute, the compressive stress is only due to the residual stress and ~300 times lower than $\sigma_{xx}$ (at the ripple transition point), i.e. negligible.

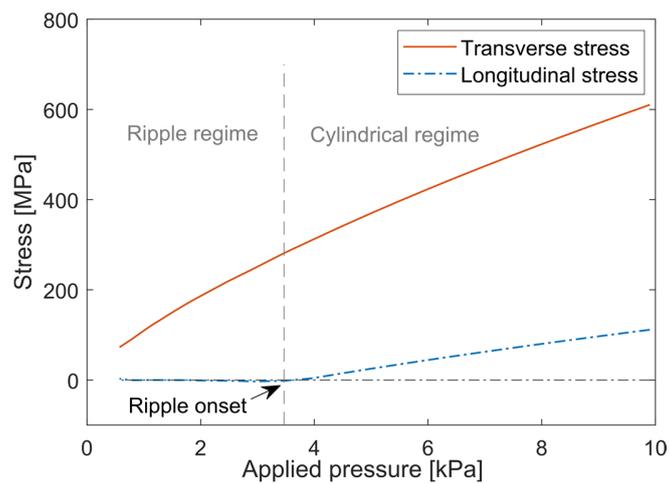

*Figure 9: Calculated transverse and longitudinal stress at the center in transverse direction and near the center in longitudinal direction, i.e. in the middle of a ripple where the curvature is zero, as a function of the applied pressure in the ripple and cylindrical regimes, revealing that the longitudinal stress stays close to zero in the rippled regime.*

Accordingly, this configuration is close to uniaxial tension, i.e. there is only a non-zero stress in the transverse direction while the deformation in the longitudinal direction is governed by free contraction. Therefore Eq. (16) and (17) reduce to:

$$\epsilon_{yy} = -\nu \frac{1}{E} \sigma_{xx}. \tag{18}$$

$$\epsilon_{xx} = \frac{1}{E} \sigma_{xx}. \tag{19}$$

Inserting Eq. (19) into Eq. (18):

$$\epsilon_{yy} = -\nu \epsilon_{xx}. \tag{20}$$

Eq. (19) and (20) provide a direct (explicit) relationship between the relevant stress and strain components, which is linear, similar to the conventional plane strain bulge test. Furthermore, unlike the plane strain bulge test, where the Young's modulus and the Poisson's ratio are coupled in the plane strain modulus ($\frac{E}{1-\nu^2}$), here $E$ and $\nu$ can be obtained separately from a single experiment.

At this point it is important to note that in the regions where the ripples have locally diminished along the longitudinal direction, e.g. point A in Figure 7, neither plane strain, nor plane stress condition holds with respect to the in-plane directions due to which Eq. (16) and (17) cannot be reduced to a simpler form to yield $E$ and $\nu$, even in a coupled form. Therefore, the analysis is best performed at the true center of the membrane (i.e. the middle in longitudinal direction), where the ripples disappear for the highest pressure, making the above-mentioned analysis valid for the largest pressure regime (another reason is that the rigid body motion is lowest at the true center as discussed above).

The validity of the uniaxial tensile state (Eq. (19) and (20)) is assessed with a virtual experiment. The stress and strain state is extracted from a node at the cross-over (zero curvature) point (denoted as point C in Figure 11) in the ripple in the center of the membrane. The Young's modulus is extracted from the gradient of the $\sigma_{xx} - \epsilon_{xx}$ plot (see Figure 10a), whereas the Poisson's ratio is extracted from the $\epsilon_{yy} - \epsilon_{xx}$ plot (Figure 10b). The Young's modulus (with a value of $154.5\ GPa$) is extracted with an error of 0.68 %, while the resulting Poisson's ratio (with a value of 0.357) reveals an error of 1.7 % with respect to the respective reference (input) values in the FE model.

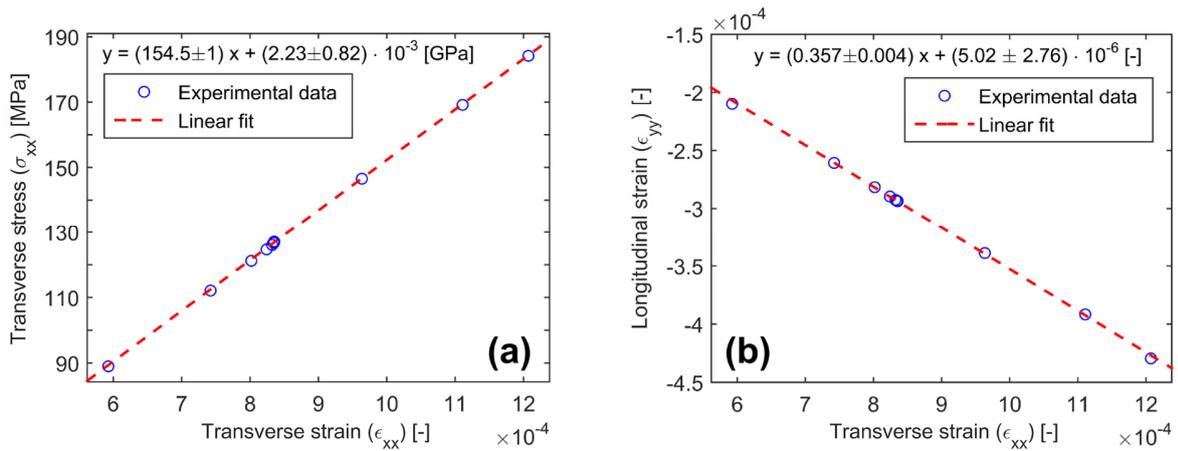

*Figure 10: Stress and strain values extracted from FE results in the rippled regime to obtain Young's modulus ($E$) and Poisson's ratio ($\nu$): **(a)** Transverse stress ($\sigma_{xx}$) vs. transverse strain ($\epsilon_{xx}$) with a linear fit to provide E, and **(b)** Longitudinal strain ($\epsilon_{yy}$) vs. transverse strain ($\epsilon_{xx}$) with a linear fit to provide $\nu$.*

Note that Eq. (15) is still valid in the rippled regime, allowing to relate the local curvature and applied pressure to the transverse stress at the crossover point. This is the case, since at the crossover point the local curvature in the longitudinal direction is almost zero, whereas $\sigma_{yy}$ is negligible w.r.t $\sigma_{xx}$ (as illustrated in Figure 9), thus the pressure applied to the face of the membrane is entirely balanced by the transverse stress. The validity of Eq. (15) in the rippled regime is numerically assessed by computing the transverse stress $\sigma_{xx}$ from $\kappa_{tt}$ (extracted from the FE simulation) and the applied pressure $P$ (known in the simulation). The resulting value was plotted against the transverse strain $\epsilon_{xx}$ extracted from the FE simulation to determine the Young's modulus, which has an error of 1.4 %.

Based on the relatively small errors in the extracted values, it is concluded that the proposed method is promising. This analysis sets the basis of the characterization methodology. In a real experiment, $\sigma_{tt}$ can be determined using Eqs. (12) - (15) from the position field in the deformed configuration. While $\epsilon_{yy}$ and $\epsilon_{xx}$ can be obtained from the 3D displacement fields using Eqs. (26) - (29), as will be shown in the next section. Furthermore, for the membranes that have not ruptured before reaching the cylindrical regime at higher pressure, this method can be used in conjunction with the plane strain equation. In that case, the same ROI analyzed in the rippled region can also be analyzed in the cylindrical plane strain state.

Practical application of DHC to the bulged buckled membranes requires highly accurate determination of 3D displacement fields as well as the position field in the reference configuration. This procedure is first tested in a virtual experiment. To this end, the displacement fields are extracted from the FE simulation in a typical ROI, from peak to valley in the longitudinal direction (marked in Figure 11a point A to B).

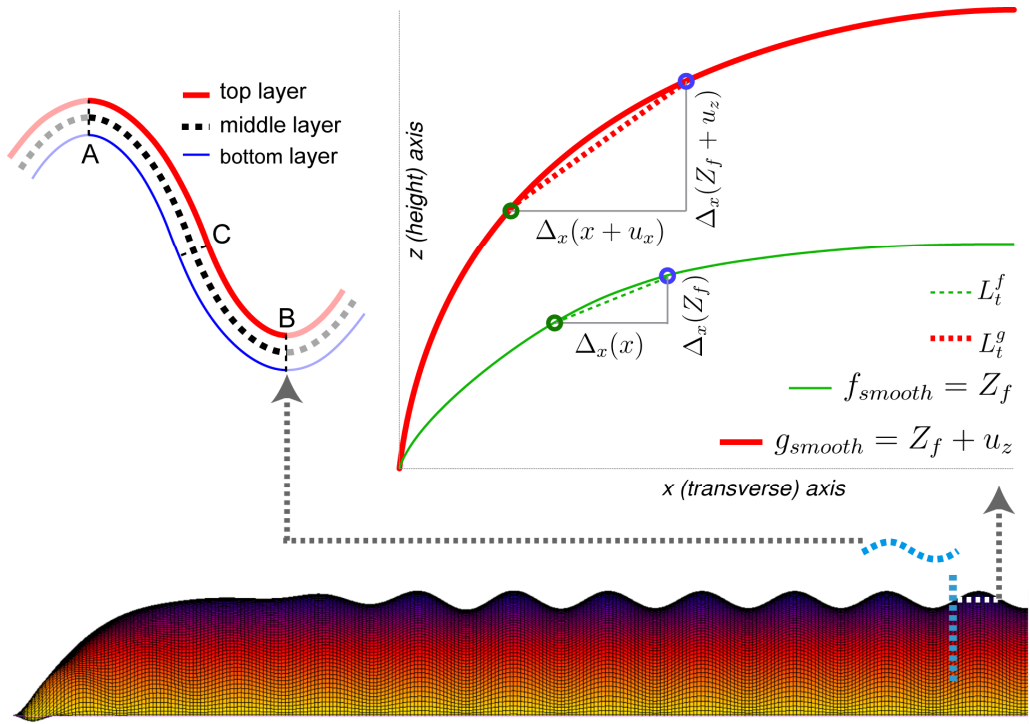

*Figure 11: A schematic illustrating: top, middle and bottom layers of membrane long the longitudinal direction **(a)**; and $L_t^g$ and $L_t^f$ used for determining the transverse membrane strain ($\epsilon_{xx}$) **(b)**; and longitudinal and transverse sections represented on FE simulation of rippled membrane **(c)**. $\Delta_x(\cdot)$ and $\Delta_y(\cdot)$ denote operators that act on the specified field to produce the field's finite difference in, respectively, the x- and y- direction.*

These displacement fields and the resulting analysis correspond to the membrane mid-plane. In DHC however, only the membrane surface (top or bottom, see Figure 11a) can be analyzed and thus only displacements at the surface are determined. This causes a discrepancy of less than 0.5 % between the transverse displacements extracted from the mid-plane and the outer surface (top/bottom) plane in the FE simulation. However, for the longitudinal displacement, due to the significant curvature in that direction, the displacement at the surface is found to be considerably different with respect to the displacement at the mid-plane. Nevertheless, the average of the longitudinal strain over a full and half ripple width (as considered here), on the top surface should be the same as on the mid-plane, since the difference in the length of a line profile from the peak to the valley of the ripple in the top surface, middle and bottom planes (Figure 11a) will be negligible, especially for small strains. This has been validated by the FE simulation. Therefore, the average longitudinal strain over half the ripple width is computed, and plotted against the transverse strain in Figure 13a, resulting in a value of Poisson's ratio with an error of only 0.3 %.

Another important observation can be made from Figure 12b, which shows that the longitudinal displacement is in the order of a few nanometers. The best optical in-plane resolution is approximately $250\ nm$. Even considering a subpixel resolution of ~0.01 pixels that can be captured with DIC techniques, the longitudinal displacement cannot be captured accurately due to the physical resolution limit. Furthermore it has been discussed in literature [26] that for large out-of-plane rotations (high curvature changes), a systematic error in the displacement fields might be introduced. To address this shortcoming, the longitudinal (in-plane) strain is disregarded and only the rigid body motion in the longitudinal direction is used, together with the height displacement field for determining the longitudinal membrane strain $\epsilon_{yy}$, see Figure 13b. The resulting Poisson's ratio still has an error limited to 0.3%. Therefore, only the rigid-body motion will be included as the displacement description in DHC to capture the longitudinal displacement field.

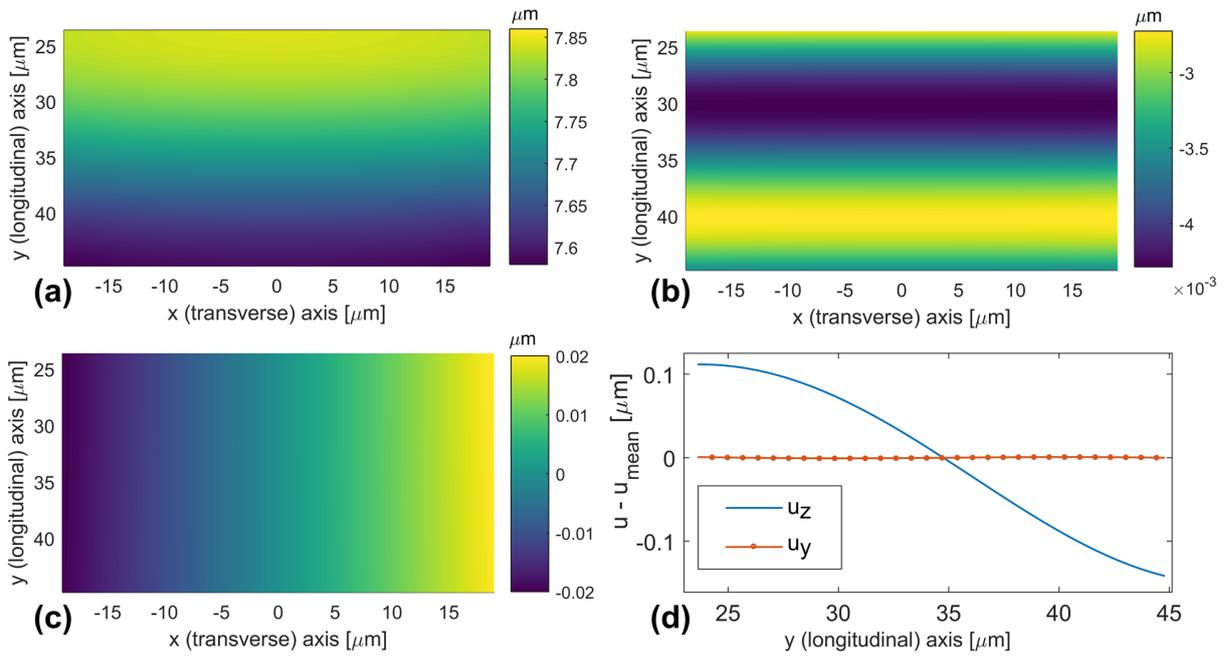

*Figure 12: Displacement fields and line-profiles extracted from FE simulations from peak to valley along the longitudinal axis in the rippled regime: **(a)** out-of-plane displacement field ($u_z$), **(b)** longitudinal displacement field ($u_y$), **(c)** transverse displacement field ($u_x$), **(d)** mean-subtracted line-profiles of $u_y$ and $u_z$ at $y = 33\ \mu m$.*

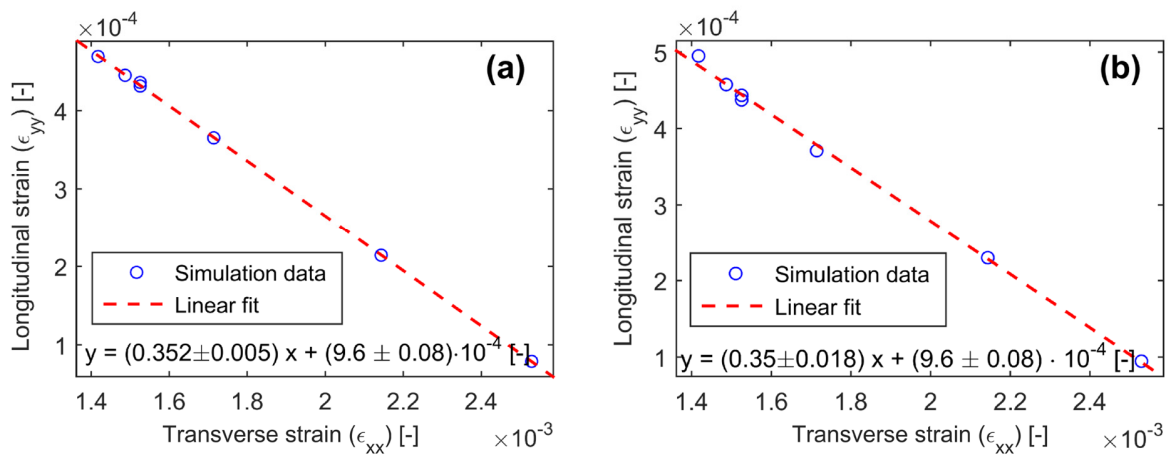
29

*Figure 13: Graphs of the longitudinal strain ($\epsilon_{yy}$) as a function of the transverse strain ($\epsilon_{xx}$) linear fits to extract the Poisson's ratio, where $\epsilon_{yy}$ is averaged over a line-profile that ranges from ripple peak to valley, for **(a)** $\epsilon_{yy}$ obtained from both $\overline{u}_y$ and $\overline{u}_z$ (from FE simulations) and **(b)** $\epsilon_{yy}$ obtained only from $\overline{u}_z$.*

# 4. Proof of principle experiment

Here, the results of a successful test serving as a proof of principle experiment to show the feasibility of application of the method in a real experiment are shown. Pressure increments of 2.5 $kPa$ were applied to the sample and kept constant while the topographical images were acquired with the profilometer.

An ROI shown in Figure 4 from the peak to valley of a ripple is selected for DHC analysis. As discussed in Ref. [8], a limited number of degrees of freedom (dofs) has to be used to capture the deformation kinematics accurately. Too few dofs restrict the kinematics while too many can cause the correlation to diverge while making it sensitive to image noise. Here, the expected order of displacements is determined from the out-of-plane and transverse displacement fields extracted from the FE simulations (Figure 12), defining the initial set of shape functions to be used for the DHC analysis. More shape functions are subsequently added to optimize the correlation until the residual fields cannot be further minimized. The longitudinal displacement only consists of a rigid-body term.

$$u_x = \lambda_1 x^0 y^0 + \lambda_2 x^0 y^1 + \lambda_3 x^1 y^0 + \lambda_4 x^1 y^1 \qquad (21)$$

$$u_y = \lambda_5 x^0 y^0 \tag{22}$$

$$u_z = \lambda_6 x^0 y^0 + \lambda_7 x^0 y^1 + \lambda_8 x^1 y^0 + \lambda_9 x^1 y^1 + \lambda_{10} x^2 y^0 + \lambda_{11} x^2 y^1 + \lambda_{12} x^2 y^2 \tag{23}$$

$$+ \lambda_{13} x^0 y^2 + \lambda_{14} x^1 y^2 + \lambda_{15} x^1 y^3 + \lambda_{16} x^0 y^3 + \lambda_{17} x^0 y^4 + \lambda_{18} x^0 y^5$$

$$+ \lambda_{19} x^0 y^6 + \lambda_{20} x^0 y^7 + \lambda_{21} x^4 y^3 + \lambda_{22} x^1 y^4 + \lambda_{23} x^1 y^5 + \lambda_{24} x^1 y^6$$

$$+ \lambda_{25} x^1 y^8 + \lambda_{26} x^3 y^8 + \lambda_{27} x^5 y^8 + \lambda_{28} x^6 y^8 + \lambda_{29} x^2 y^9$$

The number and order of shape functions needed to capture the out-of-plane displacement is significantly higher than those required for the in-plane (transverse) displacement field, see Eq.(21) and (23). This is expected since the predicted (from FE simulations) displacement field in the height direction is considerably more complex than the in-plane (transverse) displacement field, see Figure 12. This is not a problem since the optimization of the dofs related to out-of-plane shape functions is more robust than the in-plane shape functions, since the out-of-plane displacement field directly affects the image residual, whereas, the in-plane displacement field affects the image residual through the (noisy) image gradient (see Eq. (7)).

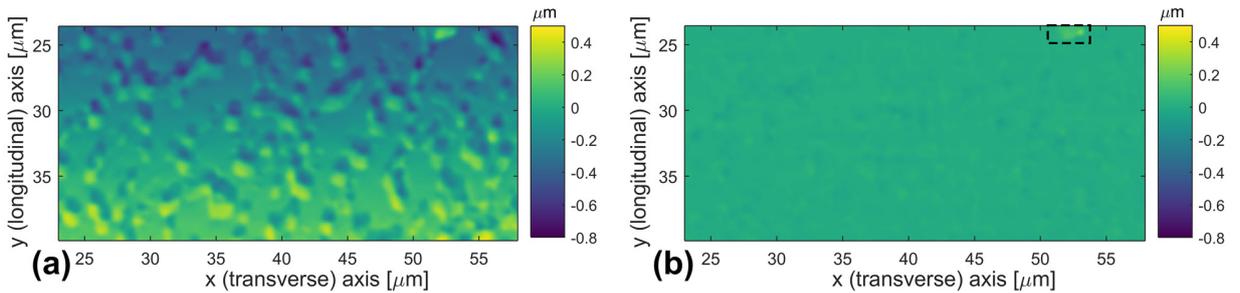

Figure 14: Residual fields for selected region of interest: **(a)** before correlation, i.e. $f(\vec{x}) - g(\vec{x})$ and **(b)** after correlation, i.e. $f(\vec{x}) - g(\vec{x} + u_x(\vec{x}) + u_y(\vec{x})) + u_z(\vec{x})$. Note the region in the black rectangle in (b), which contains an irreproducible feature, was masked from the correlation.

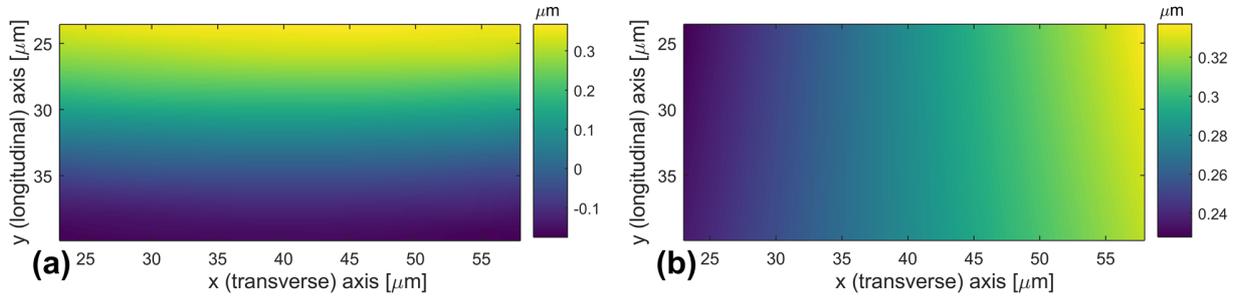

*Figure 15: Displacement fields for selected region of interest obtained by DHC: **(a)** height displacement field ($u_z$) and **(b)** transverse displacement field ($u_x$).*

The quality of the correlation in DHC is typically assessed by analyzing the residual fields. Figure 14a, and Figure 14b show the residual before and after correlation, respectively. The signature of the pattern is vaguely visible after convergence, due to the increasing rotation of the membrane during bulging, resulting in an altered viewing angle. However, the residual field has a very low RMS value indicating that the correlation was successful. Moreover, a good qualitative agreement can be seen between the shape of the displacement fields extracted from the simulations (Figure 12) and the displacement fields obtained from DHC analysis (Figure 15), confirming that the deformation kinematics has been adequately captured.

Subsequently, the required membrane strain $\epsilon_{xx}$ and $\epsilon_{yy}$ are calculated from the obtained 3D displacement fields. This involves determination of strain with respect to a non-flat (rippled) initial configuration ($f$). The stretch ratios and principal logarithmic strains resulting from the deformation are given by:

$$\epsilon_{xx} = \ln(\lambda_t) \qquad (24)$$

$$\epsilon_{yy} = \ln(\lambda_l) \qquad (25)$$

where $\lambda_t$ and $\lambda_l$ are the stretch ratios ($\lambda = \frac{current\ length}{original\ length}$) in the transverse and longitudinal directions, respectively, and given by:

$$\lambda_t = \frac{L_t^g}{L_t^f} \qquad (26)$$

$$\lambda_l = \frac{L_l^g}{L_l^f} \qquad (27)$$

$L_t^g$ and $L_t^f$ are denoted in Figure 11b and defined as:

$$L_t^g = \sqrt{\left(\Delta_x(x + u_x)\right)^2 + \left(\Delta_x(Z_f + u_z)\right)^2} \qquad (28)$$

$$L_t^f = \sqrt{\left(\Delta_x(x)\right)^2 + \left(\Delta_x(Z_f)\right)^2} \qquad (29)$$

where $Z_f$ is the height map of the reference image (i.e. $f(\vec{x})$), while $\Delta_x(\cdot)$ denotes an operator that acts on a field to produce the finite difference of the field in the x-direction. Since the reference surface is expected to be non-smooth due to the applied pattern and its intrinsic surface roughness, the height map and resulting displacement fields will be non-smooth. A surface polynomial fit of the reference image is used to smoothen this field. Similarly, for determining the surface normal and curvature values using Eq. (13), a smoothed position field ($Z_f + u_z$) is adopted instead of the deformed configuration (topographical image) $g(\vec{x})$. Likewise, $L_l^g$ and $L_l^f$ are given by:

$$L_l^g = \sqrt{\left(\Delta_y(y + u_y)\right)^2 + \left(\Delta_y(Z_f + u_z)\right)^2} \qquad (30)$$

$$L_l^f = \sqrt{\left(\Delta_y(y)\right)^2 + \left(\Delta_y(Z_f)\right)^2} \qquad (31)$$

where $\Delta_y(\cdot)$ is the operator that produces the finite difference of the specified field in the $y$-direction. Subsequently the transverse stress is determined using Eqs. (12) - (15). The transverse stress $\sigma_{xx}$ is plotted again the transverse strain $\epsilon_{xx}$ in Figure 16a resulting in a Young's modulus of $211 \pm 8\ GPa$. In order to obtain the Poisson's ratio, the mean longitudinal strain $\bar{\epsilon}_{yy}$ (instead of $\epsilon_{ll}$, as discussed in the previous section) is plotted against the transverse strain $\epsilon_{xx}$ resulting in a value of $0.36 \pm 0.12$. These values are in adequate agreement with the predicted (volume averaged) Young's modulus and Poisson's ratio of respectively $217\ GPa$ and 0.35, respectively. Therefore, the proposed methodology can be applied to obtain the elastic properties from a buckled membrane.

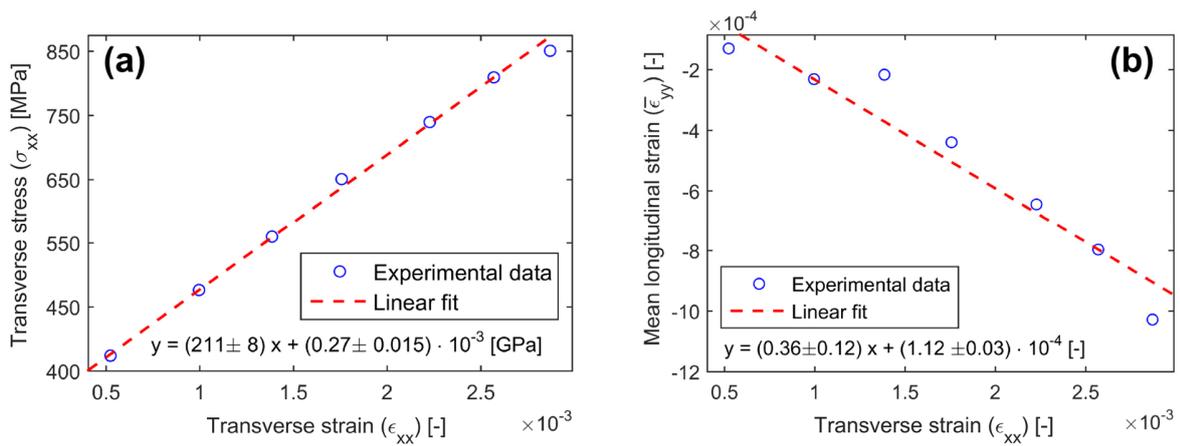

*Figure 16: Stress and strains obtained from DHC results plotted to determine Young's modulus $E$ and Poisson's ratio $\nu$: **(a)** transverse stress ($\sigma_{xx}$) vs. transverse strain ($\epsilon_{xx}$) with a linear fit to obtain the Young's modulus ($E$) and **(b)** mean longitudinal strain ($\bar{\epsilon}_{yy}$) vs. transverse strain ($\epsilon_{xx}$) with a linear fit to obtain the Poisson's ratio ($\nu$).*

# 5. Conclusions and Recommendations

It is well known from literature that conventional bulge testing, which is frequently used to characterize freestanding membranes, does not apply to the particular class of buckled membranes. In this paper, the conventional bulge test methodology has been extended to characterize the elastic properties of buckled membranes. This has been achieved by developing a validated FE based numerical model, which captures the complex mechanics of the (pressure-loaded) buckled membrane. A recently developed DHC approach has been employed to obtain local, complex 3D displacement (strain) and curvature fields.

Interestingly, the virtual experiments revealed that, embedded in the seemingly complex ripple configuration, a simplified state of uniaxial stress exists in the transverse direction. This implies that the transverse stress can be directly related to the transverse strain, yielding Young's modulus. Poisson's ratio can be extracted directly from the ratio of the longitudinal strain and the transverse strain. This is a significant advantage over the conventional bulge test theory where Young's modulus and Poisson's ratio are intrinsically measured in a coupled manner, either through the plane strain modulus or the biaxial modulus. As a result, tests on two sample geometries, typically rectangular and square shape, are needed to obtain Young's modulus and Poisson's ratio independently.

The numerical model also showed that applying DHC to measure the displacements in the longitudinal direction would require high-order shape functions and a displacement resolution that cannot be achieved with an optical system. This problem was solved by exploiting the fact that the contribution of the 'in-plane' longitudinal displacement to the longitudinal strain is negligible, allowing to obtain an accurate value of Poisson's ratio by only considering the height displacement to determine the longitudinal strain.

Proof of principle experiment clearly show that the method is applicable on real samples, even with rather narrow ripples dimensions (20-30 $\mu m$) and only few data points, as the fragile samples were very susceptible to deformation induced failure. Residual maps indicate proper convergence and the shape of the displacement fields captured with DHC adequately match the predicted displacement fields extracted from the FE simulation, confirming that the buckled membrane kinematics are being properly captured. Moreover, the resulting values of Young's modulus and Poisson's ratio are consistent with the expected values based on the material stack.

As both the stress and strains are measured under uniaxial tension, the method is not necessarily confined to the elastic regime and should work as well in the plastic regime (yielding plasticity parameters), if the specimens would plastically deform. Furthermore, by applying cyclic pressure loading, fatigue testing, as suggested in Ref. [27] for plane strain loading, can be applied with uniaxial tension. Using a feedback loop to maintain a constant membrane stress, as suggested in Ref. [28], a creep test in uniaxial tension can be performed.

# Acknowledgments

This work was supported by the Vidi funding of J.H. (project number 12966) within the Netherlands Organization for Scientific Research (NWO). The authors would also like to greatly thank Johan Klootwijk from Philips Research for proving the samples, Jan Neggers for sharing his digital image correlation code, Roel Donders for upgrading the bulge test setup and help with experiments and Marc van Maris for technical support in the laboratory.